\documentclass[notitlepage,aps,pra,amsmath,amsfonts,amssymb,floatfix,superscriptaddress,10pt,reprint]{revtex4-2}
\usepackage{color}
\usepackage{braket}
\usepackage{mathtools}
\usepackage{dcolumn} 
\usepackage{graphicx}
\usepackage{hyperref}

\begin{document}
\title{Jordan-Wigner Transformation for the Description of Strong Correlation in Fermionic Systems}

\author{Thomas M. Henderson}
\email[Author to whom correspondence should be addressed: ]{thomas.henderson@rice.edu}
\affiliation{Department of Chemistry, Rice University, Houston, TX 77005-1892, USA}
\affiliation{Department of Physics and Astronomy, Rice University, Houston, TX 77005-1892, USA}

\author{Guo P. Chen}
\affiliation{Department of Chemistry, Rice University, Houston, TX 77005-1892, USA}

\author{Gustavo E. Scuseria}
\affiliation{Department of Chemistry, Rice University, Houston, TX 77005-1892, USA}
\affiliation{Department of Physics and Astronomy, Rice University, Houston, TX 77005-1892, USA}
\date{\today}

\begin{abstract}

Seniority is a useful way of organizing Hilbert space for strongly correlated systems.  The exact zero-seniority wave function, doubly-occupied configuration interaction (DOCI), provides accurate results (given the right orbitals) for many strongly correlated electronic systems but has a combinatorial computational cost.  In many cases, pair coupled cluster doubles provides a polynomial-cost approximation that closely reproduces the energies of DOCI, but it breaks down in some cases and, as shown herein, it does not provide particularly good density matrices.  In this work, we demonstrate that by using the Jordan-Wigner transformation to turn the seniority zero problem back into a fermionic one, we can provide mean-field variational results of DOCI quality for the Hubbard model and a few small molecular dissociation examples, with polynomial cost, both for the energies and for density matrices, all while being protected from collapse.  This success is rooted in the proof we provide showing that the Hartree-Fock wave function on the Jordan-Wigner--transformed Hamiltonian transforms back to variational coupled cluster doubles in the seniority zero representation, but restricted to have determinant rather than permanent amplitude coefficients, without compromising its overall accuracy.
\end{abstract}

\maketitle

\section{Introduction}
Over the past several years, the concept of seniority has emerged as a powerful tool for the organization of Hilbert space in strongly correlated electronic systems.\cite{Bytautas2011,Johnson2012,Chen2015,CaleroOsorio2025}

Seniority is conceptually simple: every spinorbital $\varphi_P$ is paired with a counterpart $\varphi_{\bar{P}}$, and the seniority of a determinant is equal to the number of paired spinorbitals which between them are occupied by a single electron.  Seniority depends on the choice of orbitals and pairing schemes, but with suitable choices, we observe that, in many cases, strongly correlated systems are dominated by the seniority-zero sector of the Hilbert space, where all electrons are paired.

The exact wave function within the seniority zero sector of Hilbert space is known as doubly-occupied configuration interaction\cite{Allen1962, Smith1965, Weinhold1967, Veillard1967,Couty1997,Kollmar2003,Bytautas2011,Alcoba2024} (DOCI) and has also been referred to by many other names in the literature.  It has combinatorial scaling: the number of determinants in DOCI is roughly the square root of the number of determinants in the full configuration interaction (FCI).  While DOCI calculations are feasible for systems noticeably larger than those for which FCI is available, ultimately they are simply too expensive.

Fortunately, we have low-scaling alternatives.  Perhaps most familiar is the antisymmetrized product of one-reference orbital geminals (AP1roG),\cite{Limacher2013,Boguslawski2014} also known as pair coupled cluster doubles (pCCD).\cite{Stein2014,Henderson2014b}  This method has $\mathcal{O}(N^3)$ scaling, though of course optimizing the orbitals and transforming them scales as $\mathcal{O}(N^5)$.  Quite surprisingly, we empirically observe that with optimized orbitals (and, to a lesser extent, with non-optimized orbitals), pCCD closely tracks DOCI in many strongly-correlated systems.\cite{Limacher2013,Stein2014,Henderson2014b}

However, we wish to note a few shortcomings of pCCD here.  First, in strongly-correlated systems with attractive interactions, pCCD can break down and overcorrelate badly.\cite{Dukelsky2003,Henderson2014} Second, even in strongly-correlated repulsive systems, correlations can become strong enough that higher-order excitations are necessary.  Third, adding dynamic correlation atop pCCD is not entirely straightforward.\cite{Henderson2014b,Limacher2014b,Garza2015,Boguslawski2015,Garza2015b,Tecmer2022,Nowak2023}  Additionally, the fact that pCCD is nonvariational means that one must always treat the orbital optimization -- a variational energy minimization -- with a degree of care.  Finally, we note that pCCD better predicts the energy than it does the density matrices,\cite{Henderson2015} meaning that pCCD properties are less accurate than are pCCD energies in general.

Of the alternatives to pCCD, perhaps the most well-developed is the use of Richardson-Gaudin (RG) states as varitional ansatze for the zero-seniority problem.\cite{debaerdemacker2017,Johnson2020,Fecteau2022,Johnson2024,Johnson2025} Here, one finds eigenstates $\ket{\Psi_{RG}}$ of a model Hamiltonian
\begin{equation}
H_{RG} = \sum_p \epsilon_p \, \left(c_{P}^\dagger \, c_P + c_{\bar{P}}^\dagger \, c_{\bar{P}}\right) - G \, \sum_{pq} c_{P}^\dagger \, c_{\bar{P}}^\dagger \, c_{\bar{Q}} \, c_Q
\label{Eqn:HRG}
\end{equation}
where our notational convention is that pairs $(P,\bar{P})$ are indexed by $p$.  One then minimizes the expectation value of the physical Hamiltonian with respect to the parameters $\{\epsilon,G\}$ of $H_{RG}$.  These methods are intriguing and show great promise, but are somewhat specialized at the moment.

Inspired by recent work in which we used Hartree-Fock in combination with Jordan-Wigner (JW) transformation to study spin systems,\cite{Henderson2022,Chen2023,Henderson2024a,Henderson2024c,GhassemiTabrizi2025} here we apply these same techniques to the zero-seniority sector for fermionic systems.  We will see that one obtains results close to those of DOCI, at least for the Hubbard Hamiltonian and selected small molecules.

\section{The (Local) Seniority-Conserving Hamiltonian}
As we have discussed before,\cite{Henderson2015} one can always convert a physical fermionic Hamiltonian into one which conserves local seniority (i.e. it conserves the seniority of each pair $p$) through a relatively simple procedure.  We summarize it here for clarity.  We will assume for simplicity that the Hamiltonian is written in a restricted basis and that the pair $(P,\bar{P})$ corresponds to $(p_\uparrow,p_\downarrow)$, i.e. the $\uparrow$ and $\downarrow$ spinorbitals corresponding to the spatial orbital indexed by $p$.  The derivation can easily be extended to the case of unrestricted pairing where $(P,\bar{P})$ correspond to $\uparrow$ and $\downarrow$ spinorbitals with different spatial components, and even to general pairing  between noncollinear spinorbitals -- but note that in these cases, additional terms appear.

Suppose that one has a Hamiltonian
\begin{align}
H &= E_0 + \sum_{pq}  h_{pq} \, \sum_{\sigma} c_{p_\sigma}^\dagger \, c_{q_\sigma}
\\
 &+ \frac{1}{2} \, \sum_{pqrs} v_{pqrs} \, \sum_{\sigma\sigma^\prime} c_{p_\sigma}^\dagger \, c_{q_{\sigma^\prime}}^\dagger \, c_{s_{\sigma^\prime}} \, c_{r_\sigma}
\nonumber
\end{align}
where the summation over orbital indices $pqrs$ runs over spatial orbitals, the summations over spin indices $\sigma$ and $\sigma^\prime$ run over $\{\uparrow,\downarrow\}$, and $v$ is a non-antisymmetrized two-electron integral in Dirac notation.

To extract the local seniority conserving part, we simply find all ways of pairing spatial orbital indices.  For the one-electron part of the Hamiltonian, this means setting $p=q$; for the two-electron part we may have $p=q$ and $r=s$, or $pq=rs$ with $p \ne q$, or $pq=sr$ with $p \ne q$.

Enforcing these constraints gives us\cite{Henderson2015}
\begin{subequations}
\label{Eqn:SeniorityConservingHamiltonian}
\begin{align}
H &\to H_{\delta \Omega = 0}
\\
 &= E_0 + \sum_p h_{pp} \, N_p + \sum_{pq} L_{pq} \, P_p^\dagger \, P_q
\\
 &+ \frac{1}{4} \, \sum_{p \ne q} \left(2 \, J_{pq} - K_{pq}\right) \, N_p \, N_q
\nonumber
\\
 &- \sum_{p \ne q} K_{pq} \, \vec{S}_p \cdot \vec{S}_q,
\nonumber
\end{align}
\end{subequations}
where the operators are
\begin{subequations}
\begin{align}
N_p &= c_{p_\uparrow}^\dagger \, c_{p_\uparrow} + c_{p_\downarrow}^\dagger \, c_{p_\downarrow},
\\
P_p^\dagger &= c_{p_\uparrow}^\dagger \, c_{p_\downarrow}^\dagger,
\\
P_p &= c_{p_\downarrow} \, c_{p_\uparrow},
\\
S_p^x &= \frac{1}{2} \, \left(c_{p_\uparrow}^\dagger \, c_{p_\downarrow} + c_{p_\downarrow}^\dagger \, c_{p_\uparrow}\right),
\\
S_p^y &= \frac{1}{2\, \mathrm{i}} \, \left(c_{p_\uparrow}^\dagger \, c_{p_\downarrow} - c_{p_\downarrow}^\dagger \, c_{p_\uparrow}\right),
\\
S_p^z &= \frac{1}{2} \, \left(c_{p_\uparrow}^\dagger \, c_{p_\uparrow} - c_{p_\downarrow}^\dagger \, c_{p_\downarrow}\right),
\end{align}
\end{subequations}
and the Coulomb, exchange, and pairing two-electron integrals are
\begin{subequations}
\begin{align}
J_{pq} &= v_{pqpq},
\\
K_{pq} &= v_{pqqp},
\\
L_{pq} &= v_{ppqq}.
\end{align}
\end{subequations}
This Hamiltonian is the portion of $H$ that conserves the seniority of every level when seniority is defined in terms of a closed-shell restricted pairing scheme.

In the particular case that every level is seniority zero (i.e. doubly occupied or empty), the levels are spin zero, and we can disregard the spin term entirely.  Additionally, for this particular case we have $N_p^2 = 4 \, P_p^\dagger \, P_p = 2 \, N_p$, not as operator identities but in terms of the way they act on the desired states.  Since we will consider only seniority zero for the remainder of this manuscript, we will discard the spin term.

\section{Jordan-Wigner Transformation}
The Jordan-Wigner transformation\cite{Jordan1928} was introduced at the very dawn of quantum mechanics.  It relies on the observation that there is a similarity between spin 1/2 and spinless fermions: both have two states ($\ket{\uparrow}$ and $\ket{\downarrow}$ for spins, $\ket{0}$ and $\ket{1}$ for fermions) interconverted by nilpotent raising and lowering operators (for spins) or creation and annihilation operators (for fermions).  The chief difference is that spin operators on different sites commute, while fermion operators for different fermions anticommute.  We can introduce JW string operators to reconcile this distinction.

In the notation we have employed in this manuscript, and with the realization that spin operators $S^+$, $S^-$, and $S^z$ obey $\mathfrak{su}(2)$ commutation relations just as do $P^\dagger$, $P$, and $N$, we write
\begin{subequations}
\begin{align}
P_p^\dagger &\mapsto a_p^\dagger \, \phi_p^\dagger,
\\
P_p &\mapsto a_p \, \phi_p,
\\
N_p &\mapsto 2 \, a_p^\dagger \, a_p \coloneq 2 \, n_p,
\end{align}
\end{subequations}
where $a_p^\dagger$ and $a_p$ are creation and annihilation operators for new spinless fermions, and the JW string is
\begin{equation}
\phi_p = \mathrm{e}^{\mathrm{i} \, \pi \, \sum_{k < p} n_k} = \prod_{k < p} \left(1 - 2 \, n_k\right)
\end{equation}
and can be generalized to what we refer to as the extended JW (EJW) string,\cite{Wang1990,Henderson2024a}
\begin{equation}
\phi_p = \mathrm{e}^{\mathrm{i} \, \sum_k \theta_{kp} \, n_k}
\end{equation}
where the angles $\theta$ satisfy
\begin{subequations}
\label{Eqn:EJWConstraints}
\begin{align}
&\theta_{kk} = 0,
\\
&|\theta_{kp} - \theta_{pk}| = \pi.
\end{align}
\end{subequations}
In other words, the matrix $\boldsymbol{\theta}$ has zeros along the diagonal; we can take, e.g., the lower triangle to be independent, and the upper triangle is its transpose but with the addition of $\pi$ to every element.  As we have shown in Ref.~\onlinecite{Henderson2024a}, the HF solution of the JW-transformed Hamiltonian $H_\mathrm{JW}$ depends on the order in which we write the orbitals even though the ordering is irrelevant when $H_\mathrm{JW}$ is solved exactly. This labeling dependence is eliminated with the EJW-transformed Hamiltonian when the parameters $\boldsymbol{\theta}$ are optimized together with the HF state.  In other words, EJW-HF is permutationally invariant in any number of dimensions and with Hamiltonians of any type, without increasing the polynomial scaling (except for a change in the prefactor), despite the strong dependence of JW-HF on the arrangement of orbitals.

The idea in this manuscript is straightforward: we take the seniority-conserving Hamiltonian $H_{\delta\Omega=0}$ of Eqn. \eqref{Eqn:SeniorityConservingHamiltonian} and do an EJW transformation, resulting in
\begin{align}
H_{\delta\Omega = 0} \mapsto H_\mathrm{EJW} &= E_0 + \sum_p h_{pp} \, n_p 
\\
&+ \sum_{pq} L_{pq} \, a_p^\dagger \, \phi_p^\dagger \, \phi_q \, a_q
\nonumber
\\
&+ \frac{1}{4} \, \sum_{p \ne q} \left(2 \, J_{pq} - K_{pq}\right) \, n_p \, n_q.
\nonumber
\end{align}
If we solve this Hamiltonian exactly (i.e. we do a full configuration interaction calculation), we obtain the same result as we would get from the DOCI solution of $H_{\delta\Omega=0}$.  Solving it with HF, and variationally optimizing the $\boldsymbol{\theta}$ parameters, gives us what we refer to as EJW-HF,\cite{Henderson2024a,GhassemiTabrizi2025} a method with polynomial scaling variationally bounded by DOCI.

To see that the scaling is polynomial, note that the string-free terms in $H_\mathrm{EJW}$ yield typical Hartree-Fock--style expectation values so can be evaluated in $\mathcal{O}(N^4)$ cost.  For the $L_{pq}$ term, we can move the EJW strings $\phi_p^\dagger \, \phi_q$ to the right through the annihilation operator $a_q$ at the cost of a few phase factors.  We thus get an energy contribution of the form
\begin{equation}
\Delta E_{pq} \sim L_{pq} \bra{\Phi} a_p^\dagger \, a_q \, \phi_{pq} \ket{\Phi}
\end{equation}
where $\ket{\Phi}$ is the Hartree-Fock state and $\phi_{pq}$ is a one-body exponential obtained from the strings.  Via the Thouless theorem,\cite{Thouless1960} the EJW strings act on one single determinant to produce another:
\begin{equation}
\phi_{pq} \ket{\Phi} = \ket{\Phi_{pq}}.
\end{equation}
Then the energy contribution is of the form
\begin{equation}
\Delta E_{pq} \sim L_{pq} \bra{\Phi} a_p^\dagger \, a_q \ket{\Phi_{pq}} = L_{pq} \, \braket{\Phi|\Phi_{pq}} \, \rho_{qp}^{\Phi \, \Phi_{pq}}
\end{equation}
where
\begin{equation}
\rho_{sr}^{\Phi\Psi} = \frac{\bra{\Phi} a_r^\dagger \, a_s \ket{\Psi}}{\braket{\Phi|\Psi}}
\end{equation}
is the $sr$ entry of the transition density matrix between states $\bra{\Phi}$ and $\ket{\Psi}$; the entire transition density matrix can be evaluated in $\mathcal{O}(N^3)$ cost using a nonorthogonal generalization of Wick's theorem,\cite{Balian1969} to be implemented in a numerically robust way.\cite{Chen2023}  There are $\mathcal{O}(N^2)$ such terms, so the energy evaluation is $\mathcal{O}(N^5)$.  The gradient evaluation is similar.  Detailed expressions for the energy and gradients with respect to the Thouless parameters defining $\ket{\Phi}$ and the EJW angles $\theta_{pq}$ can be found in Ref. \onlinecite{GhassemiTabrizi2025}.

The $\mathcal{O}(N^5)$ cost of EJW-HF is greater than the $\mathcal{O}(N^3)$ cost of pCCD, but since the orbital optimization of either method is at least $\mathcal{O}(N^5)$ itself, the additional cost is not too high.  We will not consider the simpler JW-HF case here, simply because it introduces an unphysical dependence on the order in which we label the levels created by the operators $a_p^\dagger$; while JW-HF is simpler and has fewer parameters to optimize, its scaling is no better than that of EJW-HF.

\section{The JW Transformation of the JW-HF Wave Function
\label{Sec:JWHFWaveFunction}}
The JW-HF wave function (let us leave the EJW amplitudes aside for now, to simplify the discussion) is, in its native fermionic representation,
\begin{equation}
\ket{\Phi}_F = \mathrm{e}^{\sum Z_i^a \, a_a^\dagger \, a_i} \ket{0}_F
\label{Eqn:JWHFPsi}
\end{equation}
where $\ket{0}_F$ is some reference fermionic determinant, with occupied levels $i$, $j$, $k$, \ldots and virtual levels $a$, $b$, $c$, \ldots.  What does this wave function become when we transform it back to the $\mathfrak{su}(2)$ representation?

We can easily write this down. The reference determinant $\ket{0}_F$ transforms into some reference product state which we denote by $\ket{0}_S$; generally $\ket{0}_F$ is a single determinant in the basis in which we wrote the transformation (not in the final HF basis) and $\ket{0}_S$ correspondingly occupies the pair levels $i$, $j$, $k$, \ldots and leaves the pair levels $a$, $b$, $c$, \ldots unoccupied.  Where $\ket{0}_F$ is annihilated by the operators $a_i^\dagger$ and $a_a$, the corresponding product state $\ket{0}_S$ is annihilated by $P_i^\dagger$ and $P_a$.  The JW transformation is its own inverse, so we end up with
\begin{equation}
\ket{\Phi}_F \mapsto \mathrm{e}^{\sum Z_i^a \, P_a^\dagger \, \tilde{\phi}_a^\dagger \, \tilde{\phi}_i \, P_i} \, \ket{0}_S = \ket{\Phi}_S.
\end{equation}
The JW strings here are Hermitian (because we are using regular JW strings instead of extended strings) and are given by
\begin{equation}
\tilde{\phi}_p = \prod_{k < p} \left(1 - N_k\right).
\end{equation}

We sketch the idea here.  Fermionic determinants with operators in lexical order each map to an $\mathfrak{su}$(2) product state without a sign.  For example, 
\begin{equation}
a_1^\dagger \, a_2^\dagger \ket{-}_F \mapsto P_1^\dagger \, P_2^\dagger \, \left(1 - N_1\right) \ket{-}_S = P_1^\dagger \, P_2^\dagger \ket{-}_S
\end{equation}
where the kets $\ket{-}_F$ and $\ket{-}_S$ are the physical vacua in the two representations.  Note that
\begin{equation}
a_2^\dagger \, a_1^\dagger \ket{-}_F \mapsto P_2^\dagger \, \left(1 - N_1\right)  P_1^\dagger \, \ket{-}_S = -P_1^\dagger \, P_2^\dagger \ket{-}_S
\end{equation}
since $P_1^\dagger$ occupies level $1$ so that $N_1 \, P_1^\dagger \ket{-}_S = 2 \, P_1^\dagger \, \ket{-}_S$.  Of course this had to be true since
\begin{equation}
a_1^\dagger \, a_2^\dagger \ket{-}_F = -a_2^\dagger \, a_1^\dagger \, \ket{-}_F;
\end{equation}
the JW strings, as we expect, give us the correct signs.

This being the case, the mapping of a general wave function is straightforward:
\begin{equation}
\ket{\Psi}_F = \sum c_\mu \, \ket{\Phi_\mu}_F \mapsto \sum c_\mu \, \ket{\Phi_\mu}_S = \ket{\Psi}_S
\end{equation}
where $\ket{\Phi_\mu}_F$ is a fermionic determinant with creation operators in lexical order and $\ket{\Phi_\mu}_S$ is the corresponding $\mathfrak{su}(2)$ product state.

\begin{figure*}[t]
\includegraphics[width=0.48\textwidth]{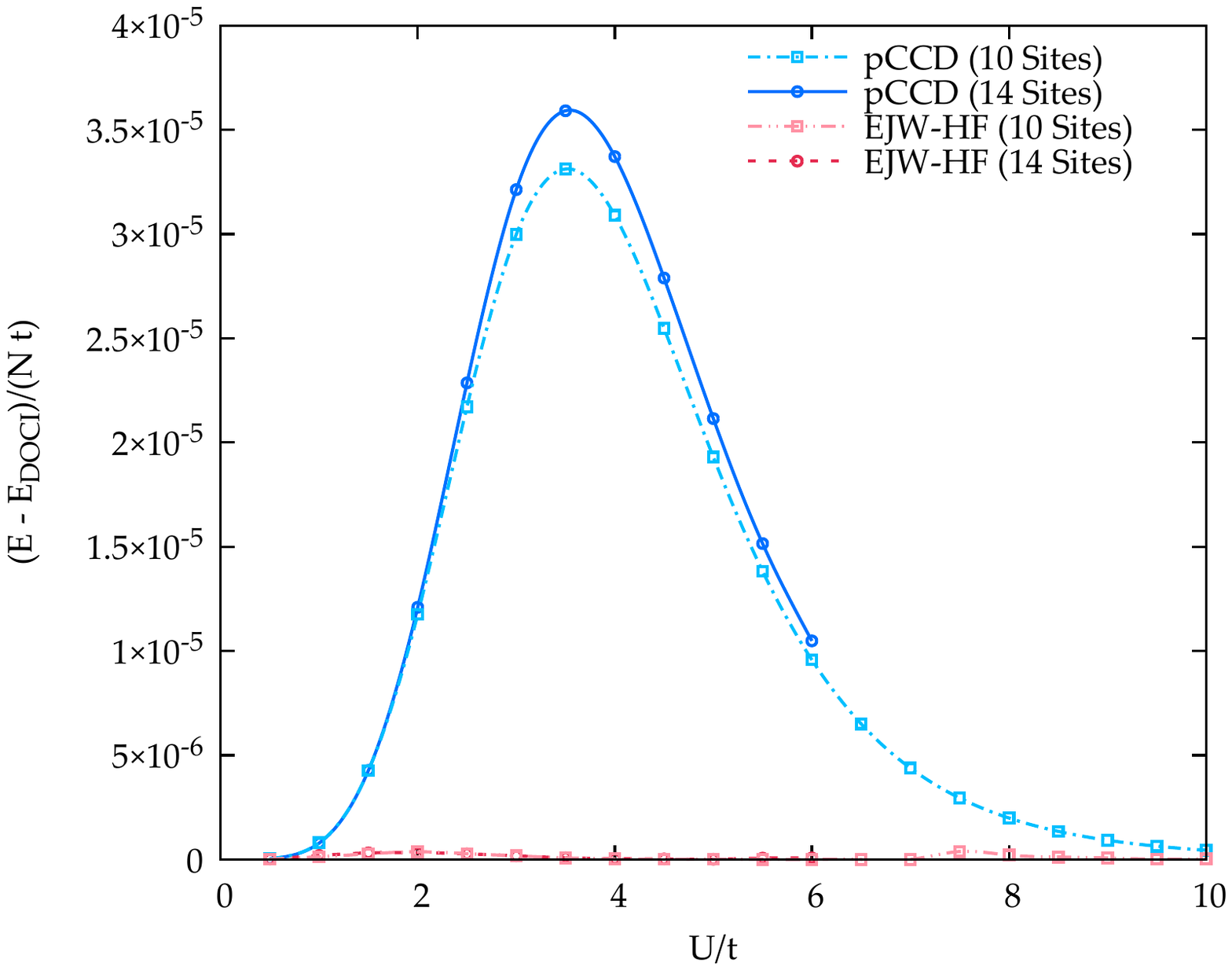}\hfill\includegraphics[width=0.48\textwidth]{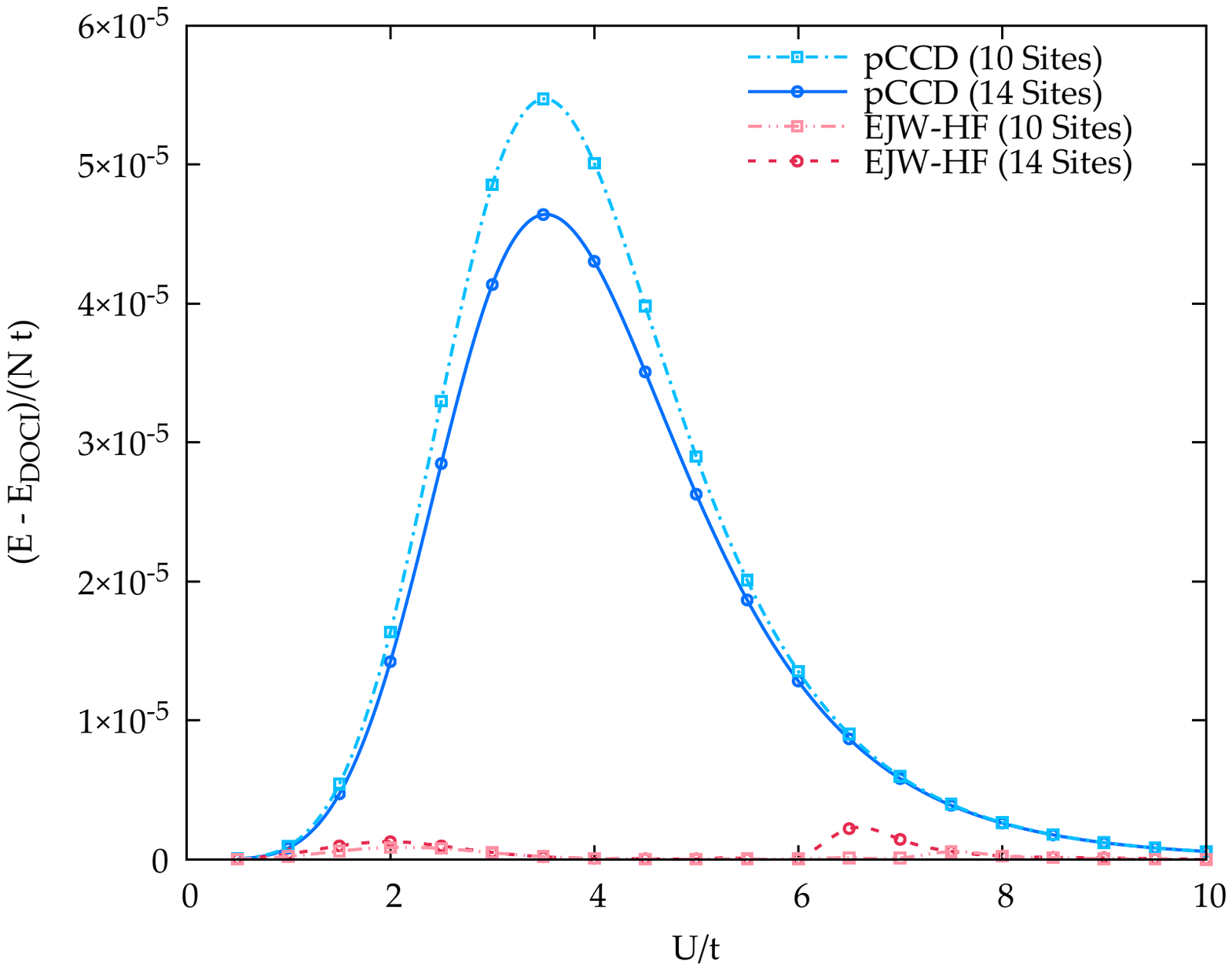}
\caption{Errors per electron with respect to DOCI in the 1D half-filled repulsive Hubbard model.  Left panel: OBC.  Right panel: PBC.  Note that we have had numerical difficulties with 14 sites with OBC so have truncated the plot at $U/t = 6$.
\label{Fig:RepulsiveHubbard}}
\end{figure*}

To understand the structure of the JW-HF wave function, we therefore need to consider the coefficients $c_\mu$ in the fermionic case (i.e. those of Eqn. \eqref{Eqn:JWHFPsi}).  They are determinants:
\begin{align}
\ket{\Phi}_F &= \ket{0}_F + \sum_{ia} Z_i^a \, \ket{\Phi_i^a}_F
+ \sum_{\substack{i<j\\a<b}} Z_{[i}^a \, Z_{j]}^b \, \ket{\Phi_{ij}^{ab}}_F
\nonumber
\\
 &+ \sum_{\substack{i<j<k\\a<b<c}} Z_{[i}^a \, Z_j^b \, Z_{k]}^c \, \ket{\Phi_{ijk}^{abc}}_F
+ \ldots
\end{align}
Here the notation $Z_{[i}^a \, Z_j^b \, Z_{k]}^c$, for example, means the original term plus signed permutations on $ijk$, while $\ket{\Phi_{ij}^{ab}}_F$, for instance, is a fermionic determinant with indices in lexical order but where we have excited levels $i$ and $j$ to levels $a$ and $b$.

Of course this means that
\begin{align}
\ket{\Phi}_S &= \ket{0}_S + \sum_{ia} Z_i^a \, \ket{\Phi_i^a}_S
+ \sum_{\substack{i<j\\a<b}} Z_{[i}^a \, Z_{j]}^b \, \ket{\Phi_{ij}^{ab}}_S
\nonumber
\\
 &+ \sum_{\substack{i<j<k\\a<b<c}} Z_{[i}^a \, Z_j^b \, Z_{k]}^c \, \ket{\Phi_{ijk}^{abc}}_S
+ \ldots
\label{Eqn:JWHFinSU2}
\end{align}
This is almost of pCCD form, but where in the EJW case the coefficients are determinants, in pCCD they are permanents instead:
\begin{subequations}
\begin{align}
\ket{\mathrm{pCCD}} &= \mathrm{e}^{\sum t_i^a \, P_a^\dagger \, P_i} \, \ket{0}_S
\\
&=
\ket{0}_S + \sum_{ia} t_i^a \, \ket{\Phi_i^a}_S
+ \sum_{\substack{i<j\\a<b}} t_{(i}^a \, t_{j)}^b \, \ket{\Phi_{ij}^{ab}}_S
\nonumber
\\
 &+ \sum_{\substack{i<j<k\\a<b<c}} t_{(i}^a \, t_j^b \, t_{k)}^c \, \ket{\Phi_{ijk}^{abc}}_S
+ \ldots
\end{align}
\end{subequations}
where the notation $t_{(i}^a \, t_j^b \, t_{k)}^c$, for example, means the original term plus \textit{unsigned} permutations on $ijk$.  Note that, as a non-Hermitian method, the pCCD bra state is much simpler.\cite{BartlettShavitt,Henderson2014b}

Alternatively, we may understand JW-HF from the perspective of a structured
tensor decomposition of the DOCI wave function. Let $\ket{\tilde{\Phi}}_F$ and
$\ket{\tilde{\Phi}}_S$ be the normalized counterparts of $\ket{\Phi}_F$
and $\ket{\Phi}_S$, respectively. We can write
\begin{equation}
  \ket{\tilde{\Phi}}_F
  = f^\dag_1 f^\dag_2 \cdots f^\dag_n \ket{-}_F,
\end{equation}
where
\begin{equation}
  f^\dag_i = \sum_p D^p_i a^\dag_p
\end{equation}
for $1 \leq i \leq n$, and $D^\cdot_i$ are the orthonormalized coefficients
of the occupied orbitals. It is readily shown that
\begin{equation}
  \ket{\tilde{\Phi}}_F
  = \sum_{p_1 < p_2 < \cdots < p_n}
  \mathcal{D}_{p_1 p_2 \cdots p_n}\,
  a^\dag_{p_1} a^\dag_{p_2} \cdots a^\dag_{p_n} \ket{-}_F,
\end{equation}
where
\begin{equation}
  \label{eq:jwhf_amp}
  \mathcal{D}_{p_1 p_2 \cdots p_n}
  = \sum_{\sigma \in S_n} \mathrm{sgn}(\sigma)
  D^{p_1}_{\sigma(1)} D^{p_2}_{\sigma(2)}
  \cdots
  D^{p_n}_{\sigma(n)}
\end{equation}
with $S_n$ being the symmetric group of order $n$. After JW transformation,
we see that
\begin{align}
  \nonumber
  \ket{\tilde{\Phi}}_S
  &= \sum_{p_1 < p_2 < \cdots < p_n}
  \mathcal{D}_{p_1 p_2 \cdots p_n}\,
  P^\dag_{p_1} \tilde{\phi}_{p_1} P^\dag_{p_2} \tilde{\phi}_{p_1}
  \cdots P^\dag_{p_n} \tilde{\phi}_{p_n} \ket{-}_S\\
  &= \sum_{p_1 < p_2 < \cdots < p_n}
  \mathcal{D}_{p_1 p_2 \cdots p_n}\,
  P^\dag_{p_1} P^\dag_{p_2} \cdots P^\dag_{p_n} \ket{-}_S,
  \label{eq:jwhf_tensor_decomp}
\end{align}
which approximates DOCI with structured amplitudes in Eqn.~\eqref{eq:jwhf_amp}.

On the other hand, pCCD or AP1roG is a special case of the antisymmetrized
product of interacting geminals (APIG),\cite{Silver1969,Silver1970,Limacher2013} which can be written as
\begin{equation}
  \label{eq:apig}
  \ket{\mathrm{APIG}}
  = \Gamma^\dag_1 \Gamma^\dag_2 \cdots \Gamma^\dag_n \ket{-}_S,
\end{equation}
where
\begin{equation}
  \Gamma^\dagger_i = \sum_p G_i^p a^\dagger_p
\end{equation}
for $1 \leq i \leq n$; the geminal coefficients $G_i^p$ are not restricted
by any orthogonal condition, though they need to satisfy additional
conditions within pCCD.\cite{Limacher2013}
We may rewrite Eqn.~\eqref{eq:apig} in the same
form as Eqn.~\eqref{eq:jwhf_tensor_decomp}:
\begin{equation}
  \ket{\mathrm{APIG}}
  = \sum_{p_1 < p_2 < \cdots < p_n}
  \mathcal{G}_{p_1 p_2 \cdots p_n}\,
  P^\dag_{p_1} P^\dag_{p_2} \cdots P^\dag_{p_n} \ket{-}_S,
\end{equation}
where
\begin{equation}
  \mathcal{G}_{p_1 p_2 \cdots p_n}\,
  = \sum_{\sigma \in S_n}
  G^{p_1}_{\sigma(1)} G^{p_2}_{\sigma(2)}
  \cdots
  G^{p_n}_{\sigma(n)},
\end{equation}
which provides a different tensor decomposition of the DOCI amplitudes.

Once we generalize all of this to EJW, things become more complicated.  The easiest way to understand it is to think of the EJW-HF wave function as combining the JW transformation with a unitary Jastrow correlator atop JW-HF, where the correlator is
\begin{equation}
\mathrm{e}^{\mathrm{i} \, J_2} = \mathrm{e}^{\mathrm{i} \, \sum_{p <q} \theta_{pq} \, n_p \, n_q}
\end{equation}
which becomes, after inverse JW transformation
\begin{equation}
\mathrm{e}^{\mathrm{i} \, J_2} \mapsto \mathrm{e}^{\frac{1}{4} \, \mathrm{i} \, \sum_{p <q} \theta_{pq} \, N_p \, N_q}.
\end{equation}
This multiplies the coefficients of Eqn.~\eqref{Eqn:JWHFinSU2} or Eqn.~\eqref{eq:jwhf_tensor_decomp} by phase factors that depend on the levels occupied but does not change their fundamentally determinantal structure.

To summarize, EJW-HF is equivalent to something akin to a variational pCCD wave function, but one where the configuration interaction expansion coefficients are determinants instead of permanents.  Such an ansatz is a constrained form (i.e., not the most general form) of resonating valence bond.\cite{Anderson1987}  We should also note the conceptual link to Zhao and Neuscamman's amplitude determinant coupled cluster (ADCC) \cite{Zhao2016}, which gives a third form: where pCCD uses a permanent and EJW-HF uses the determinant, here one uses the determinant but with a minus sign on all entries in the lower half-triangle.

We should point out that both JW-HF and ADCC give results that depend on the ordering of the levels.  As we discuss in the appendix, this is not true of EJW-HF, provided one optimizes the amplitudes $\boldsymbol{\theta}$.  For a fuller discussion of this point, see Ref. \onlinecite{Henderson2024a}.

\section{Results}
To see how well EJW-HF performs for the seniority-conserving Hamiltonian, we will consider both the Hubbard model\cite{Hubbard1963} and a few small molecular examples.  Our protocol is straightforward: we perform an orbital-optimized DOCI calculation, yielding a seniority-conserving Hamiltonian $H_{\delta\Omega=0}$, then do an EJW transformation and solve $H_\mathrm{EJW}$ at the Hartree-Fock level, simultaneously optimizing the parameters $\boldsymbol{\theta}$ in the latter case.  Of course one could directly optimize the EJW-HF result with respect to orbital choice, but since our goal here is to see how well EJW-HF reproduces DOCI, so that a DOCI calculation is necessary in any event, we have chosen not to introduce this additional complication.  We will also do pCCD calculations on the same seniority-conserving Hamiltonian.

We should emphasize that the mapping basis optimization for EJW-HF is completely feasible since it just requires the low-order physical density matrices of the EJW-HF wave function.  Because EJW-HF is fully variational, this orbital optimization can only improve the agreement with DOCI.

\subsection{The Hubbard Model}
The Hubbard model is a lattice Hamiltonian
\begin{equation}
H = -t \, \sum_{\langle pq \rangle} \sum_\sigma \left(c_{p_\sigma}^\dagger \, c_{q_\sigma}^{} + h.c.\right) + U \, \sum_p n_{p_\uparrow} \, n_{p_\downarrow}
\end{equation}
where the summations run over lattice sites, and the notation $\langle pq \rangle$ refers to sites connected in the lattice.  The strong correlation is driven by the competition between the first term (hopping), which favors delocalized orbitals, and the second term (on-site repulsion), which favors localization.

We begin by considering the 1D repulsive Hubbard model, at half filling.  Results are displayed in Fig. \ref{Fig:RepulsiveHubbard}.  We see that for both open boundary conditions (OBC) and periodic boundary conditions (PBC), pCCD closely tracks DOCI with the DOCI-optimized orbitals, with maximal errors per electron on the order of $4 \times 10^{-5} \, t$.  This is already impressively accurate, but the EJW-HF  errors are 1-2 orders of magnitude smaller yet.  We have had numerical difficulties for 14 sites with OBC so truncate the plot at $U/t = 6$, already past the point of maximal error.

\begin{figure}
\includegraphics[width=\columnwidth]{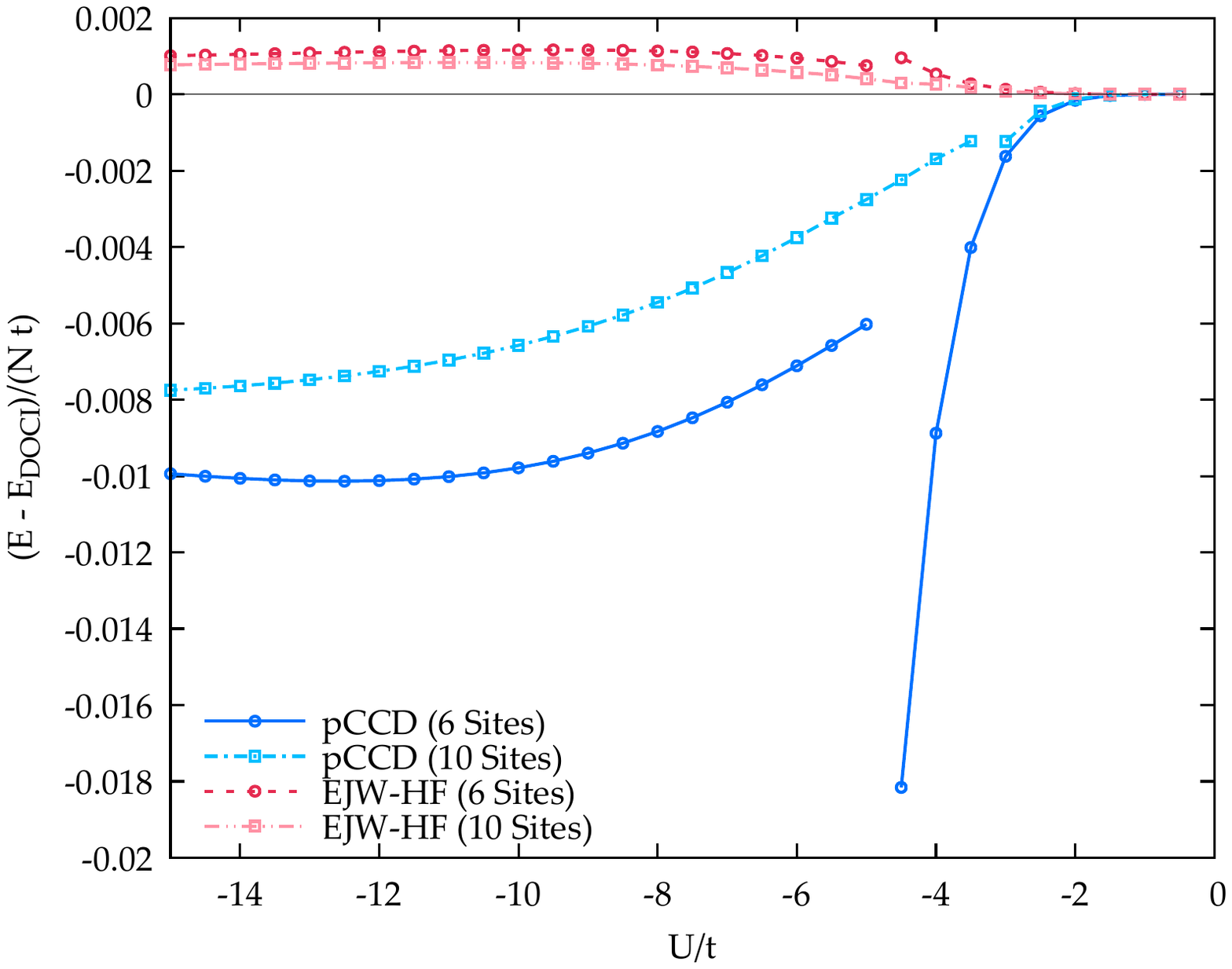}
\caption{Errors per electron with respect to DOCI in the 1D attractive Hubbard model, in PBC.  The discontinuities reflect the fact that there are two different DOCI curves that cross around $U/t \sim -4.5$ (for 6 sites) and $U/t \sim -3.5$ (for 10 sites).
\label{Fig:AttractiveHubbard}}
\end{figure}

The situation is slightly more complicated once the Hubbard model becomes attractive (see Fig. \ref{Fig:AttractiveHubbard}).  Here, we see that we have two distinct DOCI solutions.  Generally speaking, both pCCD and EJW-HF give reasonable results; EJW-HF is overall better, with errors roughly an order of magnitude smaller than pCCD.  Note also that the errors are substantially larger than in the repulsive case.  In the EJW-HF case, this can likely be resolved by allowing the EJW wave function to break number symmetry, forming an EJW-Hartree-Fock-Bogoliubov (EJW-HFB) state and then projectively restoring the broken number symmetry to form EJW-projected HFB (EJW-PHFB).\cite{Chen2023,Henderson2024c}  It is worth noting that pCCD breaks down completely for the strongly attractive pairing Hamiltonian (the large positive $G$ limit of $H_{RG}$ of Eqn. \ref{Eqn:HRG}), and while EJW-HF is poor for the strongly attractive pairing Hamiltonian, EJW-PHFB is exact in the strongly attractive limit.\cite{Henderson2024c}

\begin{figure}
\includegraphics[width=\columnwidth]{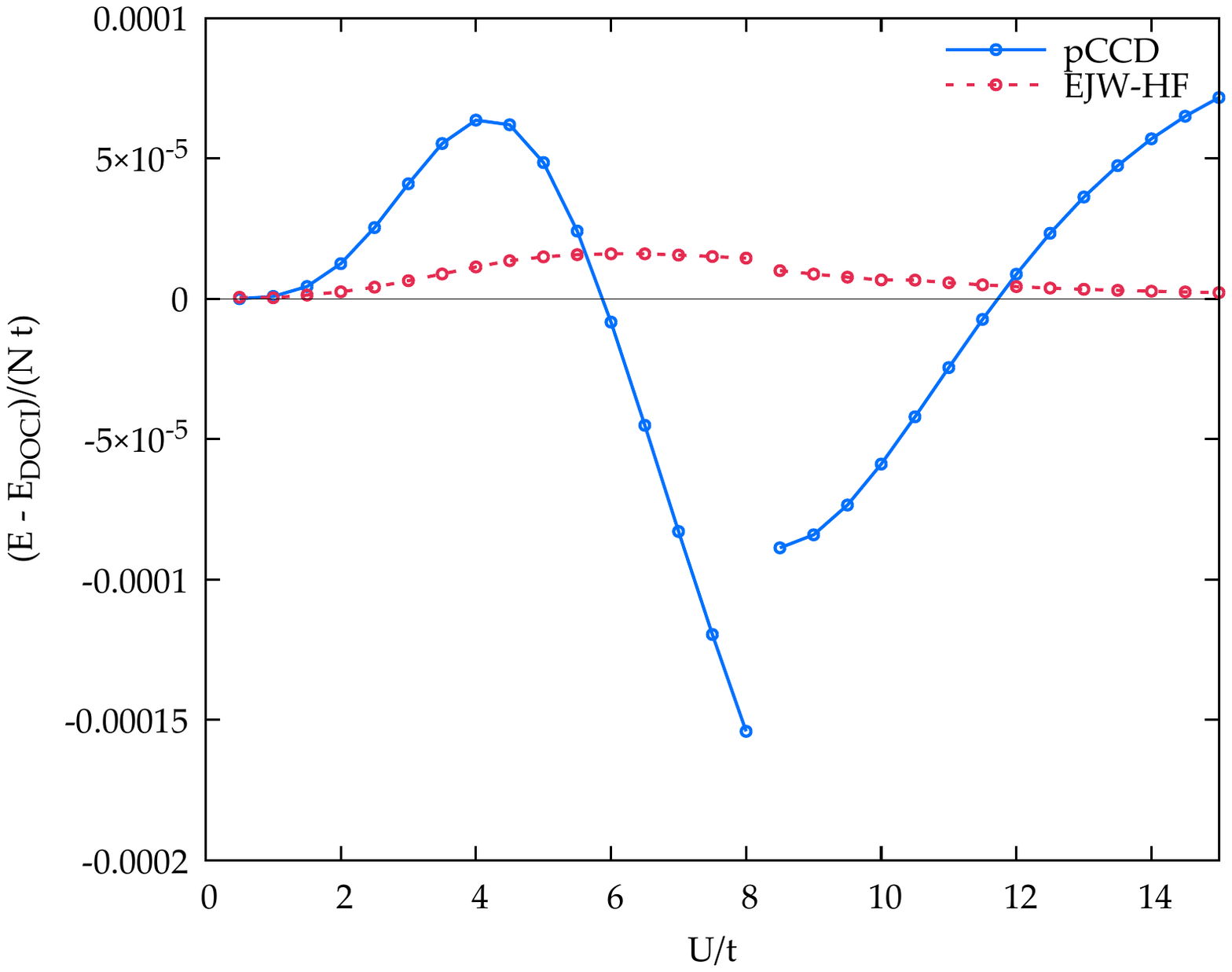}
\caption{Errors per electron with respect to DOCI in the 1D repulsive Hubbard model with 10 sites and 6 electrons, in PBC.  The discontinuities reflect the fact that there are two different DOCI curves that cross around $U/t \sim 8$.
\label{Fig:Doped1DHubbard}}
\end{figure}

Next we consider the doped Hubbard model in one dimension; see Fig. \ref{Fig:Doped1DHubbard} where we consider the 6-electron 10-site model in PBC.  Here there are two different DOCI solutions, crossing near $U/t = 8$.  We see that pCCD is again very close to DOCI, but EJW-HF is closer yet.

\begin{figure}
\includegraphics[width=\columnwidth]{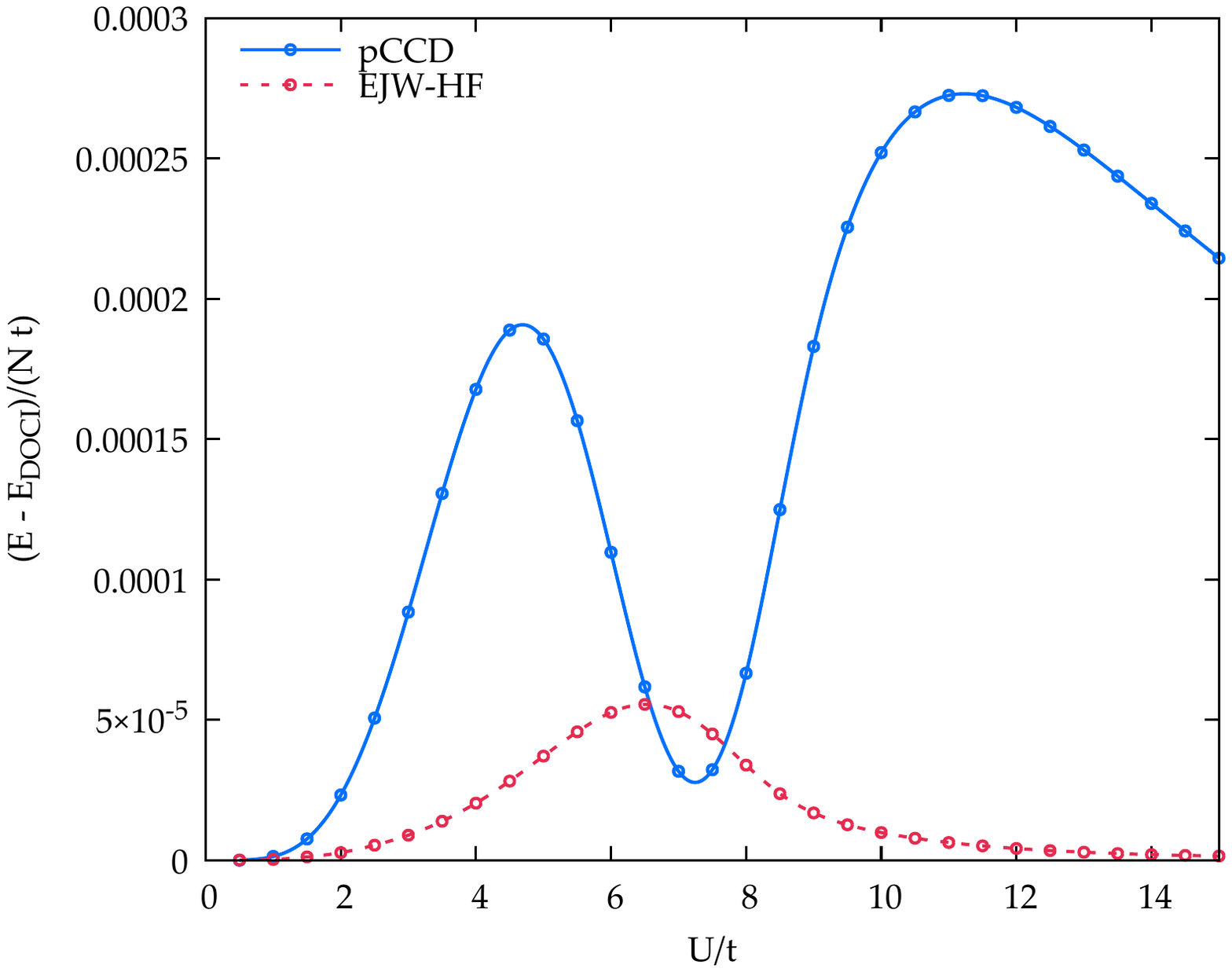}
\caption{Errors per electron with respect to DOCI in the $2 \times 4$ Hubbard model with 6 electrons, in PBC.
\label{Fig:2DHubbard}}
\end{figure}

Finally, we look at the quasi-1D $2 \times 4$ lattice with PBC, with two holes; see Fig. \ref{Fig:2DHubbard}.  While neither pCCD nor EJW-HF is as accurate as in 1D, both still reproduce DOCI quite well, and EJW-HF is once again generally better than pCCD.

\begin{figure*}
\includegraphics[width=0.3\textwidth]{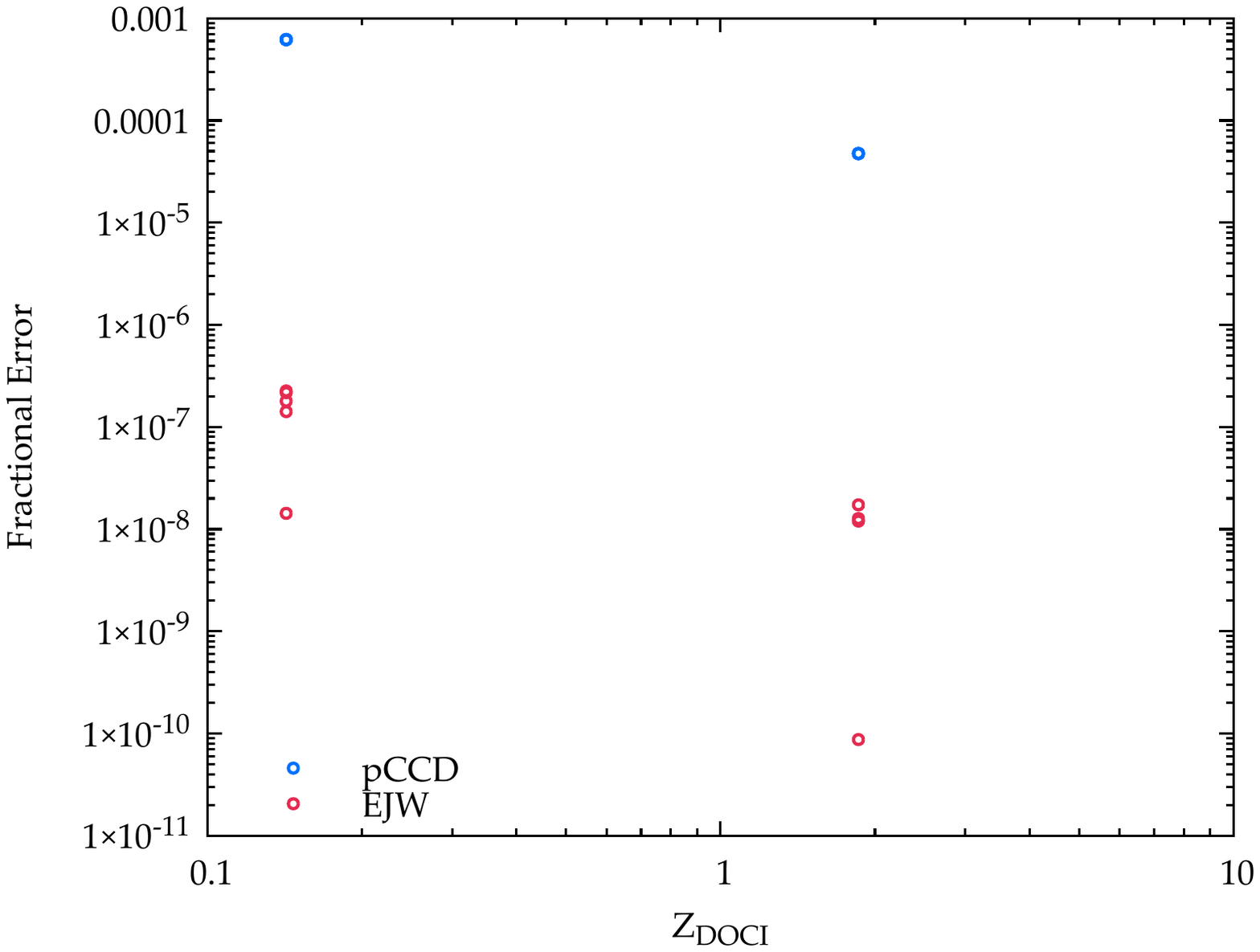}
\hfill
\includegraphics[width=0.3\textwidth]{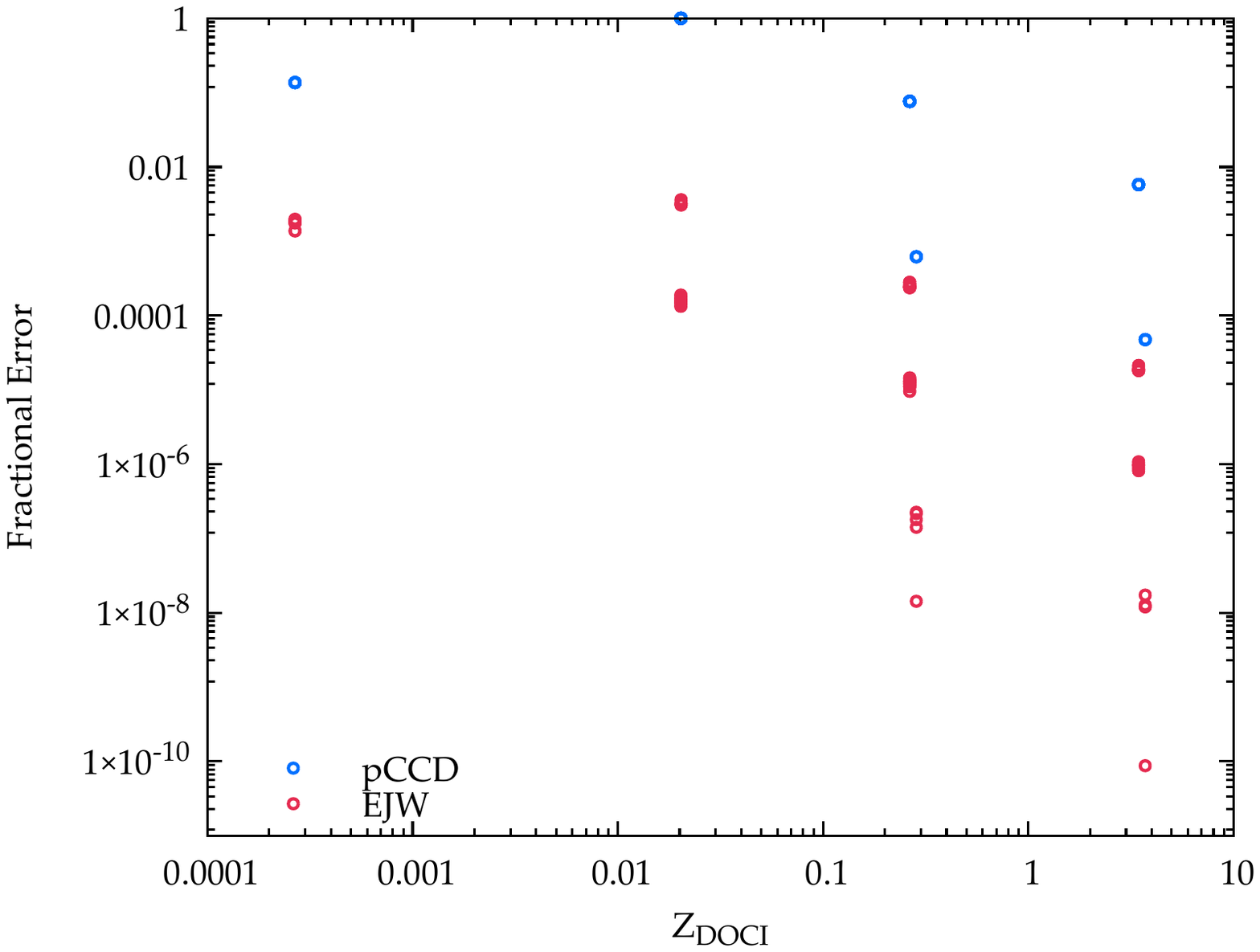}
\hfill
\includegraphics[width=0.3\textwidth]{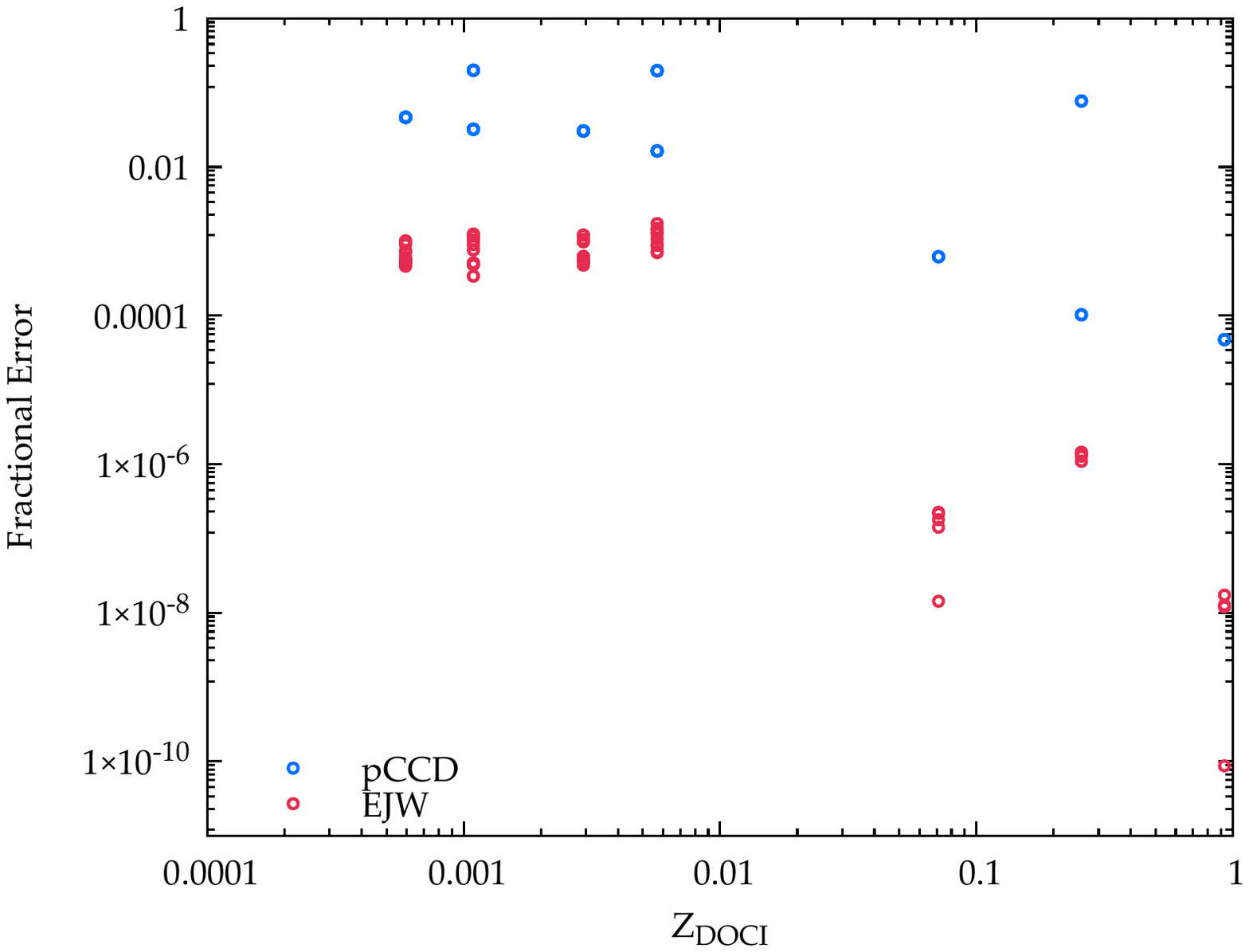}
\caption{Fractional errors in density matrix elements for the 10-site half-filled Hubbard model in PBC, at $U/t = 3.5$.  Left panel: errors in $Z^{(010)}$ (occupation numbers of DOCI orbitals).  Middle panel: Errors in $Z^{(020)}.$  Right panel: Errors in $Z^{(101)}$.
\label{Fig:DensityMatrices}}
\end{figure*}

The results so far have focused on the energies.  Generally speaking, pCCD indeed reproduces DOCI well, and there may seem limited reason to consider EJW-HF which is, after all, somewhat more expensive.  However, all we have considered are energies, and pCCD is much less capable of reproducing DOCI density matrices.\cite{Henderson2015}  It is worth recalling why this is.  Simply put, the pCCD ket $\mathrm{e}^{T_2} \ket{0}$ closely matches the DOCI ket, so that
\begin{equation}
H_{\delta \Omega=0} \ket{\mathrm{pCCD}} \approx E_\mathrm{DOCI} \, \ket{\mathrm{pCCD}}.
\end{equation}
However, the pCCD bra consists only of the reference configuration $\bra{0}$ and double excitations out of it; the pCCD bra, in other words, does \textit{not} closely resemble DOCI.  This means that for generic expectation values, pCCD can deviate significantly from DOCI.  In contrast, the EJW-HF bra and ket are adjoints of one another, so we would expect EJW-HF to more accurately reproduce DOCI density matrices and therefore properties.

To see this, we consider the particular case of the 10-site Hubbard model at half-filling, in PBC, with $U/t = 3.5$.  This is near the maximal energy error for pCCD, but the error per electron with respect to DOCI is still only approximately $3 \times 10^{-5} \, t$.  We consider three density matrices in the DOCI optimized orbital basis:
\begin{subequations}
\begin{align}
Z^{(010)}_p &= \langle N_p \rangle,
\\
Z^{(020)}_{pq} &= \langle N_p \, N_q \rangle,
\\
Z^{(101)}_{pq} &= \langle P_p^\dagger \, P_q \rangle.
\end{align}
\end{subequations}
In the EJW-HF case, we transform the operators to the fermionic representation.  For pCCD, we solve the linear response equations and define the usual coupled cluster expectation value\cite{Salter1989,BartlettShavitt,Henderson2014b}:
\begin{equation}
\braket{O}_\mathrm{pCCD} = \prescript{}{S}{\bra{0}} \, \left(1 + Z_2\right) \, \mathrm{e}^{-T_2} \, O \, \mathrm{e}^{T_2} \ket{0}_S
\end{equation}
where $Z_2$ has the same form as $T_2^\dagger$ but with different amplitudes.

Figure \ref{Fig:DensityMatrices} shows fractional errors in the density matrix elements against the DOCI density matrix elements.  That is, the $y$-axis plots 
\begin{equation}
\delta Z = \frac{|Z - Z_{\mathrm{DOCI}}|}{|Z_\mathrm{DOCI}|}.
\end{equation}
We see that the pCCD density matrix elements are quite accurate for $Z^{(010)}$ but are much worse for the two-body density matrices.  The same is true for EJW-HF, but the errors are reduced by a few orders of magnitude on average.  This implies that EJW-HF should be noticeably closer to DOCI for an arbitrary property than is pCCD.

We should note a few difficulties with the EJW density matrices.  First, there is some scatter in values that in DOCI (and pCCD) are identical, with discrepancies on the order of $10^{-5} - 10^{-8}$.  Second, $Z^{(101)}$ is weakly complex, with imaginary components on the order of $10^{-6} - 10^{-8}$.  We presume these discrepancies could be made smaller by employing tighter convergence on the wave function parameters.  Our fractional convergence threshold on the energy is $10^{-12}$ implying, for a variational method such as EJW-HF, convergence in the wave function parameters on the order of $10^{-6}$, which would seem to explain the small fluctuations in density matrix elements.  Recall that in DOCI, the different levels are all treated equivalently (and the same is true in pCCD, except that levels occupied in the reference and empty in the reference are treated differently).  In contrast, the JW string means that the levels in EJW are not treated entirely equivalently.  The optimization of the $\theta$ parameters in EJW resolves this issue \cite{Henderson2024a}, but presumably the energy is less sensitive than is the wave function.

While the small residual labeling-dependence in the EJW-HF density matrices is inconvenient, we suggest it is a reasonable price to pay for the significantly improved density matrices.

Note particularly that in pCCD, density matrix elements $Z^{(020)}_{ab} = \langle N_a \, N_b \rangle = 0$ for two levels $a$ and $b$ both empty in the pCCD reference $\ket{0}_S$.  This is because the first non-zero term in $N_a \, N_b \, \mathrm{e}^{T_2} \ket{0}_S$ is the $T_2^2$ term, so that we require quadruple and higher excitations, but the pCCD bra consists only of the reference and doubles.  This naturally means that pCCD has 100\% error for these density matrix elements.

At $U/t = 3.5$ in our 10-site example, the exact density matrix element $Z^{(020)}_{6,7}$ between the first two such virtual levels is approximately 0.020270; pCCD gives zero and EJW-HF gives 0.020266.  Approximating this density matrix element with zero may seem reasonable, but note that at $U/t = 15$ the same density matrix element is 0.538908 in DOCI and EJW-HF, but 0 in pCCD.  The pCCD density matrix element is disastrous, even though the pCCD \textit{energy} error is only $2 \times 10^{-8} \, t$ per electron.

\begin{figure}
\includegraphics[width=\columnwidth]{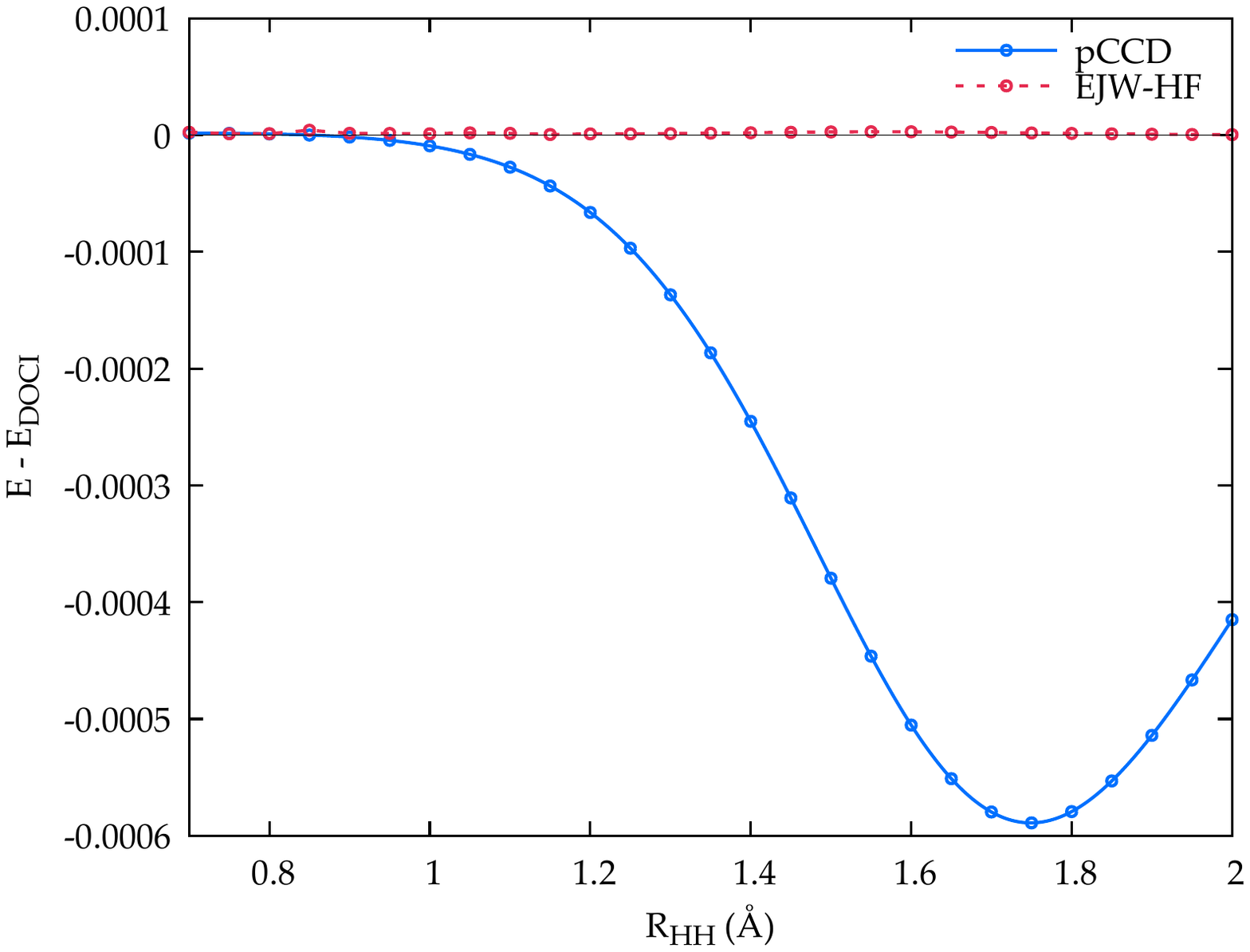}
\caption{Error vs DOCI in the minimal basis H$_8$ linear chain.
\label{Fig:H8}}
\end{figure}

\subsection{Molecular Examples}
The Hubbard model already suffices to convey our basic point: that EJW-HF provides results of DOCI quality with polynomial cost; though we have greater sensitivity to initial guess than does pCCD (not shown), our results are also generally more accurate.  Here we confirm this result for a few small molecular examples where the oo-DOCI is computationally straightforward.

We start with the minimal basis linear H$_8$ chain, with hydrogen atoms separated by $R$.  This is a sort of molecular analog of the Hubbard chain: the kinetic energy term favors delocalization while the electron-electron repulsion term favors localization.  Results are shown in Fig. \ref{Fig:H8}.  Again, pCCD very closely reproduces DOCI and EJW-HF is even better.  Since we already saw these results for the one-dimensional Hubbard model which has similar physics, this is what we expect to see.

\begin{figure}
\includegraphics[width=\columnwidth]{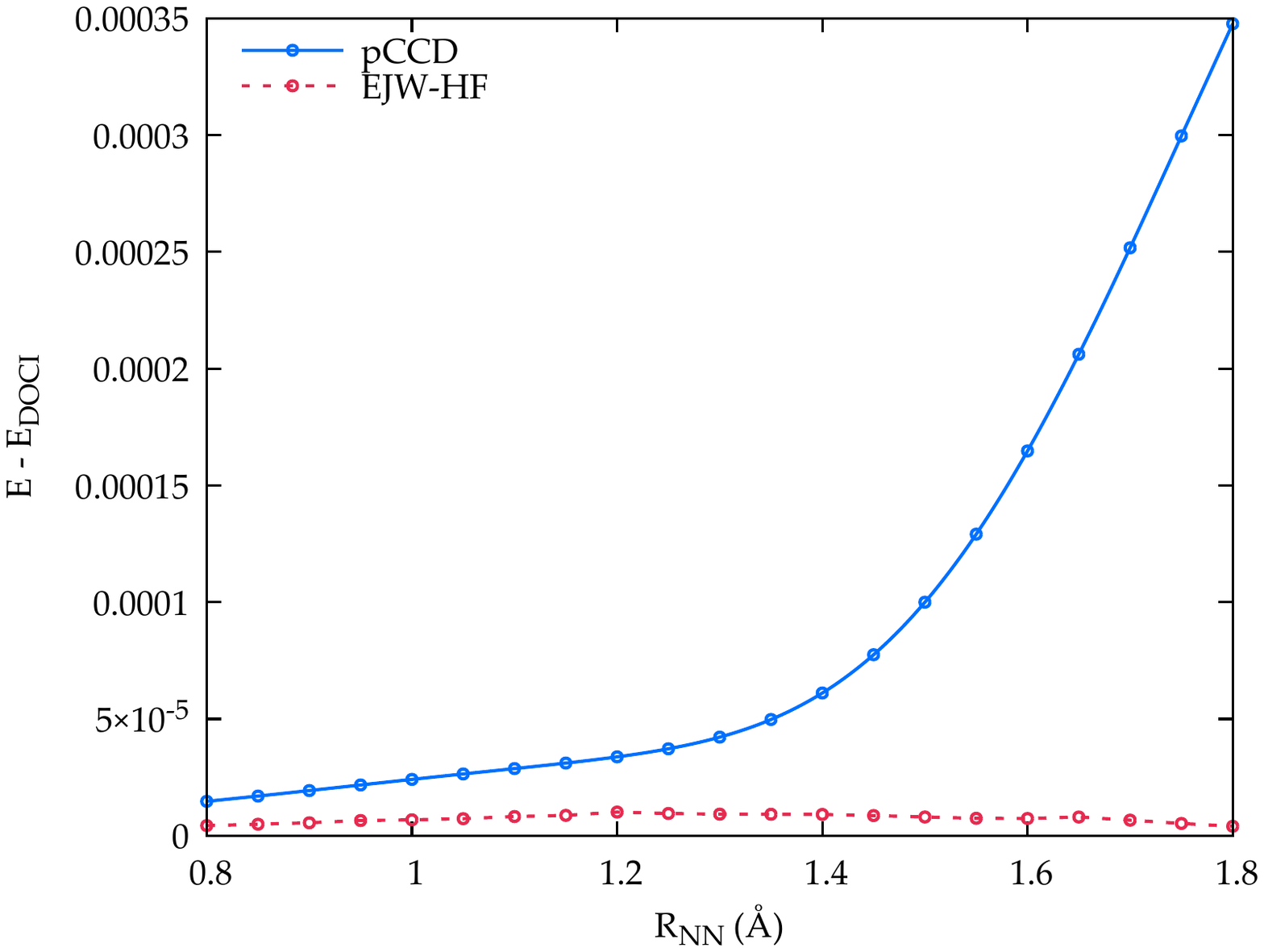}
\caption{Error vs DOCI in minimal basis N$_2$.
\label{Fig:N2}}
\end{figure}

To show that the good performance of EJW-HF is not limited to Hubbard-like systems, we finally consider minimal basis N$_2$ in Fig. \ref{Fig:N2}.  Again, pCCD behaves very well in capturing DOCI, but EJW-HF is even better, all while preserving Hermitian character and a variational bound.

\section{Discussion}
With suitably chosen orbitals, the zero-seniority sector of Hilbert space is important for the description of strong electronic correlations.  This is intuitively reasonable: the hallmark of strong correlation is that electrons cease to behave as individuals and instead behave collectively, and the simplest way to do this is to form pairs.

Unfortunately, actually using this observation to simplify the description of strongly-correlated electronic systems is not always straightforward.  Pair coupled cluster performs admirably for typical repulsive systems, such as those of chemical relevance.  This statement is truer of the energy than of the density matrices -- presumably a consequence of the inadequate left-hand state in pCCD -- and we observe additionally that pCCD can break down badly.  Moreover, as a nonvariational method, orbital optimization in the context of pCCD seems to us to be at least somewhat suspicious in that the pCCD energy functional need not be bounded from below.

Here, we seek to use a completely new alternative for the description of the zero-seniority part of the electronic Hamiltonian.  By using Jordan-Wigner transformation to convert the $\mathfrak{su}(2)$ Hamiltonian responsible for the description of the zero-seniority sector back into fermionic language, it becomes possible to treat the zero-seniority sector with all the typical tools of electronic structure theory.  We have focused here on the simplest alternative: the use of Hartree-Fock in combination with the extended Jordan-Wigner strings.  However, one could equally well use Hartree-Fock-Bogoliubov, and one could even include number projection.  Such more variationally complete treatments provide even greater accuracy without changing the computational scaling of the method.\cite{Chen2023,Henderson2024b,Henderson2024c}

We have already seen that EJW-HF is a reasonable method for strongly correlated spin systems.\cite{Henderson2022,Chen2023,Henderson2024a,Henderson2024b,Henderson2024c}  Here, we show that it also adequately captures the strong correlations in the Hubbard model (or, more precisely, it captures the strong correlations that DOCI can capture).  Because EJW-HF is a fully variational method, orbital optimization in the EJW-HF context is well justified.  And because EJW-HF is a Hermitian theory, we guarantee real energies and Hermitian density matrices, which pCCD cannot do.  Moreover, we see that the density matrices of EJW-HF appear to be more accurate than those of pCCD, though they have some remaining errors due presumably to the greater sensitivity of density matrices than energies to the variational parameters in the EJW-HF wave function.  

We should, perhaps, note an important general theme: the exact seniority zero wave function is unquestionably DOCI, but it is in general computationally intractable.  Practical methods use a product form.  Both AP1roG/pCCD\cite{Limacher2013,Boguslawski2014,Stein2014,Henderson2014b} and RG states\cite{debaerdemacker2017,Johnson2020,Fecteau2022,Johnson2024,Johnson2025} approximate DOCI as a product of geminals.  This approach seems very sensible in light of the structure of the DOCI wave function.\cite{MartinezGonzalez2025}  Here we have taken a different approach: by transforming the problem back to a fermionic representation we have written a fermionic mean-field state whose form in terms of the seniority operators $P$, $P^\dagger$, and $N$ is somewhat complicated: it is of a form similar to that of variational CCD but with coefficients that are determinants rather than permanents.  The lack of permutational invariance that our form implies is remedied by optimization of the $\boldsymbol{\theta}$ parameters defining the EJW strings.  One could, if necessary, envision correlating EJW-HF with configuration interaction or coupled cluster to reach even better agreement with DOCI; in light of our results here, this additional effort seems likely to be unnecessary (but see results for EJW-HF in spin lattice models\cite{Henderson2022,Henderson2024a,Henderson2024c}, where additional correlations are unquestionably required).

Regardless of which zero-seniority wave function one chooses, one must address the important question of how best to include the effects of higher seniority sectors.  These are crucial for describing dynamic correlation and are at times also important for strong correlations (for example, DOCI for N$_2$ dissociation still has notable errors at large bond lengths).\cite{Bytautas2011}  Moreover, we must recognize that DOCI has optimized the energy in the seniority zero sector but does not provide the correct zero seniority part of the exact FCI wave function;\cite{WahlenStrothman2018} thus, when we include higher seniorities, we must also amend the wave function in the seniority zero sector.  The ideal solution to this difficulty is not yet apparent, but it seems of pressing importance.


\begin{acknowledgments}
This work was supported by the U.S. Department of Energy, Office of Basic Energy Sciences, under Award DE-SC0019374.  G.E.S. is a Welch Foundation Chair (C-0036).
\end{acknowledgments}

\appendix
\section{Appendix: Ordering Invariance in Extended Jordan Wigner}
We showed the form of the JW-HF wavefunction in Sec. \ref{Sec:JWHFWaveFunction}.  It is not too complicated to see that JW-HF gives results that depend on how one orders the paired levels.  Here, we wish to show how once we optimize the $\boldsymbol{\theta}$ amplitudes in EJW-HF together with the Thouless parameters, this labeling dependence is eliminated.

To give a concrete example, suppose we have a 2-pair, 4-level system where JW-HF already exhibits errors (it is, of course, exact for one-pair systems) where the algebra is relatively straightforward.  The JW-HF wave function in the fermionic picture is
\begin{subequations}
\begin{align}
\ket{\Phi}_F
 &= \mathrm{e}^Z \, \ket{12}_F
\\
 &= \ket{12}_F
\nonumber
\\
 &+ Z_{2}^3 \, \ket{13}_F + Z_{2}^4 \, \ket{14}_F - Z_{1}^3 \, \ket{23}_F - Z_{1}^4 \, \ket{24}_F
\nonumber
\\
 &+ \left(Z_{1}^3 \, Z_{2}^4 - Z_{1}^4 \, Z_{2}^3\right) \ket{34}_F
\end{align}
\end{subequations}
where here we have chosen $\ket{12}_F$ as our reference and where the ket $\ket{ab}_F = c_a^\dagger \, c_b^\dagger \ket{-}_F$ in terms of the physical vacuum $\ket{-}_F$.  We have used a non-unitary Thouless representation, but of course one could use a unitary representation instead.  Now we map these back to spins.  Because the fermions are already in their canonical order, ket $\ket{ab}_F \mapsto \ket{ab}_S$ where $\ket{ab}_S = P_a^\dagger \, P_b^\dagger \ket{-}_S$.  This is as we have already discussed, and we get trivially
\begin{align}
\ket{\Phi}_S
 &= \ket{12}_S
\nonumber
\\
 &+ Z_{2}^3 \, \ket{13}_S + Z_{2}^4 \, \ket{14}_S - Z_{1}^3 \, \ket{23}_S - Z_{1}^4 \, \ket{24}_S
\nonumber
\\
 &+ \left(Z_{1}^3 \, Z_{2}^4 - Z_{1}^4 \, Z_{2}^3\right) \ket{34}_S.
\end{align}

Suppose, however, that we had sorted the orbitals not in the order $\{1,2,3,4\}$ but in the order $\{1,3,2,4\}.$  Now the fermionic ket $\ket{23}_F$, being not in lexical order according to our new ordering, maps to the spin ket $-\ket{23}_S$.  Then in this new ordering, the JW-HF wave function becomes (after transforming back to spins)
\begin{align}
\ket{\Phi^\prime}_S
 &= \ket{12}_S
\nonumber
\\
 &+ (Z^\prime)_2^3 \, \ket{13}_S + (Z^\prime)_2^4 \, \ket{14}_S + (Z^\prime)_1^3 \, \ket{23}_S - (Z^\prime)_1^4 \, \ket{24}_S
\nonumber
\\
 &+ \left((Z^\prime)_1^3 \, (Z^\prime)_2^4 - (Z^\prime)_1^4 \, (Z^\prime)_2^3\right) \ket{34}_S
\end{align}
where we have used Thouless parameters $\boldsymbol{Z}^\prime$ to emphasize that in a different ordering, one might obtain different wave function parameters.

It should be clear that no choice of $\boldsymbol{Z}^\prime$ suffices to make  $\ket{\Phi^\prime}_S = \ket{\Phi}_S$.  In other words, changing the ordering of the levels, which should be physically inconsequential, has given us a different wave function and a different result.  If we solved the problem exactly, this would not be a problem, but the HF result depends on this labeling.

Things change when we add the extended strings.  In the original ordering $\{1,2,3,4\}$, we now have
\begin{subequations}
\begin{align}
\ket{\Psi}_F
 &= \mathrm{e}^{\mathrm{i} \, J_2} \, \mathrm{e}^Z \, \ket{12}_F
\\
 &= \mathrm{e}^{\mathrm{i} \, \theta_{12}} \, \ket{12}_F
\nonumber
\\
 &+ Z_{2}^3 \, \mathrm{e}^{\mathrm{i} \, \theta_{13}} \, \ket{13}_F
  + Z_{2}^4 \, \mathrm{e}^{\mathrm{i} \, \theta_{14}} \, \ket{14}_F
\nonumber
\\
 &- Z_{1}^3 \, \mathrm{e}^{\mathrm{i} \, \theta_{23}} \, \ket{23}_F
  - Z_{1}^4 \, \mathrm{e}^{\mathrm{i} \, \theta_{24}} \, \ket{24}_F
\nonumber
\\
 &+ \left(Z_{1}^3 \, Z_{2}^4 - Z_{1}^4 \, Z_{2}^3\right) \, \mathrm{e}^{\mathrm{i} \, \theta_{34}} \, \ket{34}_F.
\end{align}
\end{subequations}
Then we transform this wave function back to spins using the JW transformation and we simply replace kets $\ket{ab}_F$ with $\ket{ab}_S$ since we have the fermionic kets ordered correctly:
\begin{align}
\ket{\Psi}_S
 &= \mathrm{e}^{\mathrm{i} \, \theta_{12}} \, \ket{12}_S
\nonumber
\\
 &+ Z_{2}^3 \, \mathrm{e}^{\mathrm{i} \, \theta_{13}} \, \ket{13}_S
  + Z_{2}^4 \, \mathrm{e}^{\mathrm{i} \, \theta_{14}} \, \ket{14}_S
\nonumber
\\
 &- Z_{1}^3 \, \mathrm{e}^{\mathrm{i} \, \theta_{23}} \, \ket{23}_S
  - Z_{1}^4 \, \mathrm{e}^{\mathrm{i} \, \theta_{24}} \, \ket{24}_S
\nonumber
\\
 &+ \left(Z_{1}^3 \, Z_{2}^4 - Z_{1}^4 \, Z_{2}^3\right) \, \mathrm{e}^{\mathrm{i} \, \theta_{34}} \, \ket{34}_S.
\end{align}

If we used a different ordering $\{1,3,2,4\}$, we would obtain
\begin{align}
\ket{\Psi^\prime}_S
 &= \mathrm{e}^{\mathrm{i} \, \theta_{12}^\prime} \, \ket{12}_S
\nonumber
\\
 &+ (Z^\prime)_2^3 \, \mathrm{e}^{\mathrm{i} \, \theta_{13}^\prime} \, \ket{13}_S
  + (Z^\prime)_2^4 \, \mathrm{e}^{\mathrm{i} \, \theta_{14}^\prime} \, \ket{14}_S
\nonumber
\\
 &+ (Z^\prime)_1^3 \, \mathrm{e}^{\mathrm{i} \, \theta_{23}^\prime} \, \ket{23}_S
  - (Z^\prime)_1^4 \, \mathrm{e}^{\mathrm{i} \, \theta_{24}^\prime} \, \ket{24}_S
\nonumber
\\
 &+ \left((Z^\prime)_1^3 \, (Z^\prime)_2^4 - (Z^\prime)_1^4 \, (Z^\prime)_2^3\right) \, \mathrm{e}^{\mathrm{i} \, \theta_{34}^\prime} \, \ket{34}_S.
\end{align}
Now, though, we can ensure that $\ket{\Psi^\prime}_S = \ket{\Psi}_S$: we set $\boldsymbol{Z}^\prime = \boldsymbol{Z}$ and $\boldsymbol{\theta}^\prime = \boldsymbol{\theta}$, except that $\theta_{23}^\prime = \theta_{23} + \pi$.  For \textit{any} reordering of the levels, we can adjust $\boldsymbol{\theta}^\prime$ and $\boldsymbol{Z}^\prime$ so that we get the same spin state regardless of ordering.

To understand the origins of this invariance, let us return to the JW string, which is the source of our labeling dependence.  The JW string is, recall,
\begin{equation}
\phi_p^\mathrm{JW} = \mathrm{e}^{\mathrm{i} \, \pi \, \sum_{k<p} n_k}.
\end{equation}
It is the restricted range of the summation that gives us the labeling dependence in JW (and also that makes JW function correctly in the first place).  But the EJW string
\begin{equation}
\phi_p^\mathrm{EJW} = \mathrm{e}^{\mathrm{i} \, \sum \theta_{kp} \, n_k}
\end{equation}
does not have this summation restriction.  We still have $\theta_{kk} = 0$ and that the lower half-triangle and upper half-triangle of $\boldsymbol{\theta}$ differ by $\pi$ (see Eqn. \ref{Eqn:EJWConstraints}), but if we change the order of two sites, we can add $\pi$ to the angle $\theta$ in both the lower and upper triangles to compensate for the change in ordering.

\bibliographystyle{apsrev4-2}
\bibliography{ref}

\end{document}